\documentstyle[prb,aps,epsf,twocolumn]{revtex}
\def\eq{\begin{equation}}
\def\be{\begin{equation}}
\def\ee{\end{equation}}
\def\eqa{\begin{eqnarray}}
\def\eea{\end{eqnarray}}

\def\s{\sigma}

\def\w{\omega}

\def\ap{\alpha}

\begin{document}
\draft
\flushbottom
\twocolumn[
\hsize\textwidth\columnwidth\hsize\csname @twocolumnfalse\endcsname
\title{Superconductivity of a Metallic Stripe Embedded in an Antiferromagnet}
\author{Yu.A.Krotov$^{(a)}$, D.-H.~Lee$^{(a)}$ and A.V. Balatsky$^{(b)}$} 
\address{(a)Department of Physics, University of California at Berkeley, Berkeley, CA 94720}
\address{(b)Theory Division, Los Alamos National Laboratory,
Los Alamos, NM 87545}
\date{\today}
\maketitle
\tightenlines
\widetext
\advance\leftskip by 57pt
\advance\rightskip by 57pt

\begin{abstract}

We study a simple model for the metallic stripes found in
$La_{1.6-x}Nd_{0.4}Sr_xCuO_4$: two chain Hubbard ladder embedded in a {\it static} antiferromagnetic environments. We consider two cases: a ``topological stripe'', for which the phase of the  Neel order parameter shifts by $\pi$  across the ladder, and a ``non-topological stripe'', for which there is no phase shift across the ladder. We perform one-loop renormalization group calculations to determine the low energy properties. We compare the results with those of the isolated ladder and show that for  small doping superconductivity is enhanced in the topological stripe, and suppressed in the non-topological one. In the topological stripe, the superconducting order parameter is a mixture of a spin singlet component with zero momentum and a spin triplet component with momentum $\pi$. We argue that this mixture is generic, and is due to the presence of a {\it new} term in the quantum Ginzburg-Landau action. Some consequences of this mixing are discussed. 
   
\end{abstract}

\vskip 1cm

]

\narrowtext
\tightenlines

Recent neutron experiments have demonstrated that for $La_{1.6-x}Nd_{0.4}Sr_xCuO_4$,
% revealed that for several doping concentration $x$, 
 doped holes segregate into an array of ``stripes'' embedded in an antiferromagnetic background.\cite{tranq1}
Moreover, it was found that for $x=0.12, 0.15, 0.20$ the systems exhibit simultaneous stripe and superconducting order.\cite{tranq2} 
According to  Tranquada {\it et al}, a consistent interpretation of the observed spin and charge incommensurability is that the stripes are the 
antiphase domain walls of the magnetic order.  A number of theoretical works have addressed issues varying from the origin of the stripe order\cite{sep}, to the effects of stripe order on superconductivity\cite{ekz,cn}.

In this paper we examine a simple model of a {\it single}  stripe. The model consists of a Hubbard ladder (where the holes reside) embedded in one of the two types of {\it static} antiferromagnetic environments.\cite{scala} 
In case (a), the magnetic order on the two sides of the ladder forces a $\pi$ 
phase shift in the Neel order parameter (Fig.1(a)). We shall subsequently refer to this 
type of stripe as being topological. In case (b), the Neel order parameter
stays in-phase across the stripe (Fig.1(b)). We refer to this case as being 
non-topological.

The low energy properties of the model are determined by performing one-loop renormalization group (RG) calculations. Comparing our results with the results of similar calculations for an isolated ladder allows us to infer the effects of a magnetic environment on the low energy properties of the two chain Hubbard ladder. We emphasize that in our view the present work sheds light on how superconductivity {\it survive} stripe ordering, but not how stripe ordering {\it trigger} superconductivity.

Our main results are summarized as follows. When the Hubbard ladder is lightly doped, we find that superconductivity is {\it enhanced} (relative to the isolated Hubbard ladder) in the topological stripe, and is {\it totally suppressed}
in the non-topological one. The superconducting order parameter in the former is a linear combination of a spin singlet component with zero-momentum, and a spin triplet component with momentum $\pi$. We argue that this mixture is generic, and is due to the presence of a new term (Eq.(\ref{gl})) in the quantum  Ginzburg-Landau action.

\begin{figure}[b]
\epsfysize=6.0cm\centerline{\epsfbox{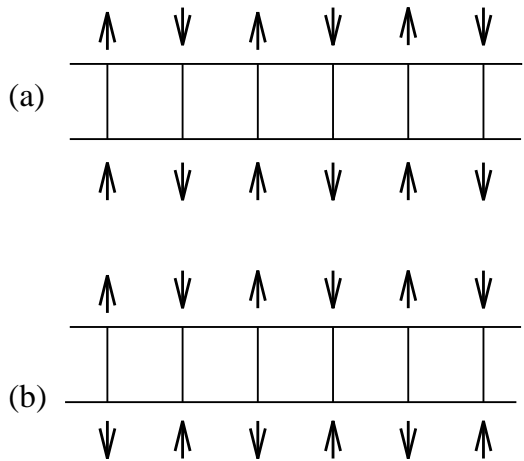}}
\vspace{20pt}
\caption{The topological (a), and non-topological (b)
stripes. The arrows represent the magnetic moments of the environment. Antiferromagnetic coupling exists between an arrow and electrons on the site next to it. To determine whether a stripe is topological, one needs to interpolate the magnetic order across the ladder.} 
\label{stripe}
\end{figure}

\vspace{0.1in}
\noindent{\bf{The model}}
\vspace{0.1in}

\rm

The model we shall study describes a metallic ``river'' (stripe) embedded in 
an insulating antiferromagnetic background.  The stripe will be modeled by a ``Hubbard ladder''(Fig.1). The coupling between the electrons in
the stripe and the magnetic moment of the background is the usual antiferromagnetic spin-spin interaction. The Hamiltonian is
\eqa
&&H=H_0+V,\nonumber \\
&&H_0=-t\sum_{i\s}\{[c^+_{i+1\s}c_{i\s}+d^+_{i+1\s}d_{i\s}
+c^+_{i\s}d_{i\s}+h.c.]\nonumber \\
&&-\mu[c^+_{i\s}c_{i\s}+d^+_{i\s}d_{i\s}]
+M\s(-1)^i[c^+_{i\s}c_{i\s}+\eta d^+_{i\s}d_{i\s}]\},\nonumber \\
&&V=\frac{U}{2}\sum_{i\s}[c^+_{i\s}c_{i\s}c^+_{i-\s}c_{i-\s}+
d^+_{i\s}d_{i\s}d^+_{i-\s}d_{i-\s}].
\label{str}
\eea
In the above, $i$ runs through the lattice sites, $\s=\pm 1$ labels the electron spin, $c$ and $d$ annihilate electrons on the two chains of the ladder, $\mu$ is the chemical
potential, $U$ is the Hubbard interaction, $M$  is the internal staggered magnetic field induced by
the background moments, and $\eta=+1(-1)$ when the stripe is topological (non-topological). 

When $M=0$ Eq.(\ref{str}) describes an isolated Hubbard ladder. The following is a brief summary of the theoretical results for this system.\cite{ladder} At half filling, the ladder is a Mott insulator  with both a spin and charge gap. Superconductivity develops when it is lightly doped. In the superconducting phase, the order parameter is the out-of-phase linear combination of the usual spin singlet order parameters in the symmetric 
and antisymmetric bands. 

When $M\ne 0$ the first Brillouin zone is halved, 
and the free-particle bands of the isolated ladder hybridize. The resulting bandstructures are shown in Fig.2.
At half filling there are four Fermi points (which pairwisely coincide) for $\eta=+1$ (Fig.2a). In contrast, an energy gap $\Delta=M$ opens up for $\eta=-1$ (Fig.2b). Doping moves the chemical potential so that for both $\eta=\pm 1$ there are four distinct 
Fermi points ($k_{F0}\pm\delta k_F$). Here 
$\delta k_F$ is the shift in Fermi momentum due to doping, and $k_{F0}$ satisfies $(M/t)^2+4\cos^2(k_{F0})=1$ for $\eta=1$, and $k_{F0}=\pm\pi/3$ for $\eta=-1$ respectively. (In the above and hereafter we will set the lattice constant to unity.)
The dispersion of the energy band that intersects 
$\mu$ at, e.g., $k_{F0}+\delta k_F$ is such that $dE_k/dk<0$ at the Fermi level. We shall define $k_{FR-}\equiv k_{F0}+\delta k_F$, and refer to the portion of energy band in its vicinity as the $(R-)$ branch. (Here $R$ designates the right group of Fermi points, and $-$ reflects the fact that $ dE_k/dk<0$.) Similar notation will be given to the Fermi momenta and energy band branches associated with the other three Fermi points.

Let $\psi_{R\pm}$ and $\psi_{L\pm}$ be the electron annihilation operators associated with $(R\pm)$ and $(L\pm)$ branches respectively. In terms of them the lattice  annihilation operator $c_{i\s}$ is given by
\eqa
c_{i\s}&=&e^{ik_{FR-}x_i}\psi_{R-\s}(x_i)+e^{ik_{FR+}x_i}\psi_{R+\s}(x_i)\nonumber \\
&+&e^{ik_{FL-}x_i}\psi_{L-\s}(x_i)+e^{ik_{FL+}x_i}\psi_{L+\s}(x_i).
\eea 
In the above $x_i$ is the coordinate of the i'th lattice site. 
\begin{figure}[b]
\epsfysize=5.2cm\centerline{\epsfbox{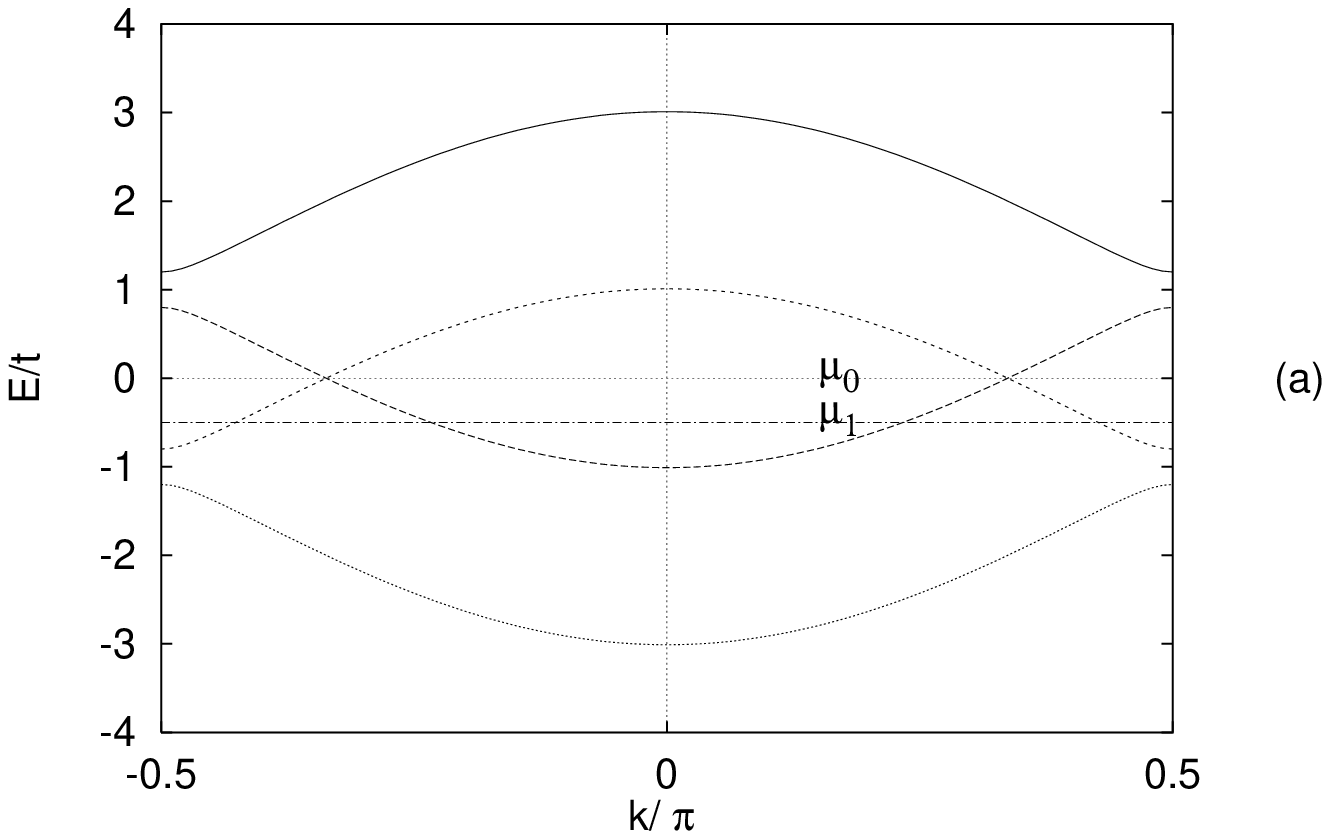}}
\vspace{20pt}
%%\caption{The bandstructure of the topological(a) and nontopological(b) stripe.} 
%%\label{band}
%%\end{figure}

%%\begin{figure}[b]
\epsfysize=5.2cm\centerline{\epsfbox{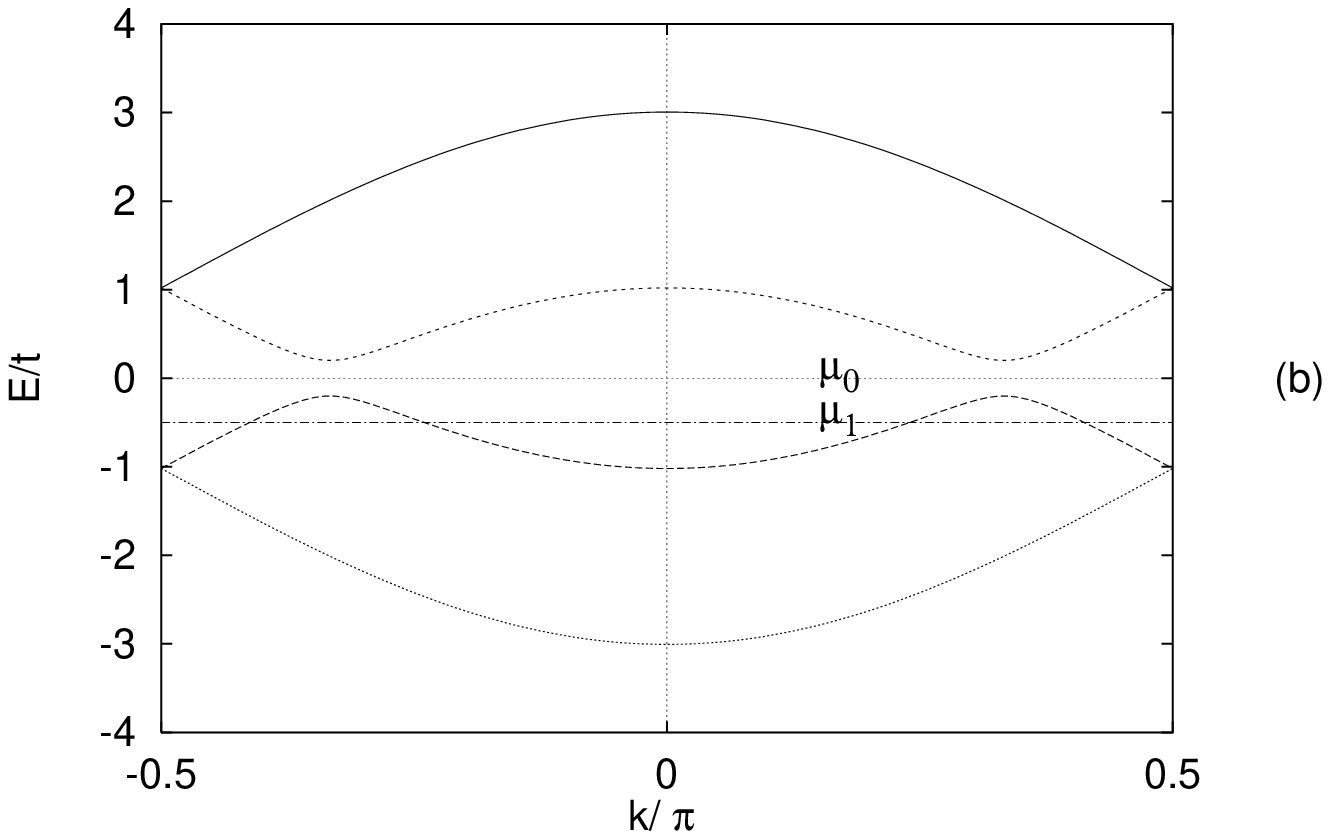}}
\vspace{20pt}
\caption{The bandstructure of the topological(a) and non-topological(b)
stripe. $\mu_0$ is the chemical potential at half filling; $\mu_1$ is the chemical potential after doping.} 
\label{band}
\end{figure}

\vspace{0.1in}
\noindent{\bf{The RG solution}}
\vspace{0.1in}

\rm

Since we are primarily interested in the low energy properties, we shall limit ourselves to a small window of the energy band near each Fermi point.
To be more precise, we assume that we can truncate the Hilbert space so that aside from the free Fermi sea, it only includes the {\it subset} of particle-hole excitations where the particle/hole states fall inside the small windows. Furthermore, within each window we shall ignore the curvature of the band. Next we project the Hubbard interaction onto the truncated Hilbert space. The result is a family of  vertex functions which describe the scattering matrix elements between states near Fermi energy. In order to simplify the calculation, we shall assume that the vertex function has  $+\leftrightarrow -$ symmetry. This requires that $v^{(i)}_F\equiv\mid dE^{(i)}_k/dk\mid$ is the same for each energy band $E^{(i)}_k$. In the rest of the paper we shall concentrate on small doping where this symmetry is approximately valid. 
%We don't expect this assumption to impact on the %qualitative physics we are about to discuss below.

For generic doping, i.e. when there is no umklapp scattering, there are
five independent vertex functions given by 
\eqa
&&\Gamma_{RRRR}^{++++}(\s_1\s_2\s_3\s_4)=
g^4_{4s}(\delta_{\s_1\s_3}\delta_{\s_2\s_4}
-\delta_{\s_1\s_4}\delta_{\s_2\s_3})\nonumber \\
&&+ g^4_{4a}\s_1\s_2(\delta_{\s_1\s_3}\delta_{\s_2\s_4}
-\delta_{\s_1\s_4}\delta_{\s_2\s_3}),\nonumber \\
&&\Gamma_{RRRR}^{-+-+}(\s_1\s_2\s_3\s_4)=(
g^2_{4s}\delta_{\s_1\s_3}\delta_{\s_2\s_4}-g^1_{4s}\delta_{\s_1\s_4}
\delta_{\s_2\s_3})\nonumber \\
&&+\s_1\s_2 (g^2_{4a}\delta_{\s_1\s_3}\delta_{\s_2\s_4}
-g^1_{4a}\delta_{\s_1\s_4}\delta_{\s_2\s_3}),\nonumber \\
&&\Gamma_{LRRL}^{++++}(\s_1\s_2\s_3\s_4)=(
g^4_{1s}\delta_{\s_1\s_3}\delta_{\s_2\s_4}-g^4_{2s}\delta_{\s_1\s_4}
\delta_{\s_2\s_3})\nonumber \\
&&+\s_1\s_2 (g^4_{1a}\delta_{\s_1\s_3}\delta_{\s_2\s_4}
-g^4_{2a}\delta_{\s_1\s_4}\delta_{\s_2\s_3}),\nonumber \\
&&\Gamma_{LRRL}^{-+-+}(\s_1\s_2\s_3\s_4)=(
g^2_{1s}\delta_{\s_1\s_3}\delta_{\s_2\s_4}-g^1_{2s}\delta_{\s_1\s_4}
\delta_{\s_2\s_3})\nonumber \\
&&+\s_1\s_2 (g^2_{1a}\delta_{\s_1\s_3}\delta_{\s_2\s_4}
-g^1_{2a}\delta_{\s_1\s_4}\delta_{\s_2\s_3}),\nonumber \\
&&\Gamma_{LRRL}^{+--+}(\s_1\s_2\s_3\s_4)=(
g^1_{1s}\delta_{\s_1\s_3}\delta_{\s_2\s_4}-g^2_{2s}\delta_{\s_1\s_4}
\delta_{\s_2\s_3})\nonumber \\
&&+\s_1\s_2 (g^1_{1a}\delta_{\s_1\s_3}\delta_{\s_2\s_4}
-g^2_{2a}\delta_{\s_1\s_4}\delta_{\s_2\s_3}).
\label{g}
\eea
All other vertices that can be generated from the above  $\Gamma$'s
by letting $R\leftrightarrow L$ and/or $+\leftrightarrow -$ have the same value. These vertex functions are characterized by 18 coupling constants $g^j_{is}$ and $g^j_{ia}$. In the above we have chosen the notation so that $i$ and $j$ reflect the 
changes in the upper ($+$ or $-$) and lower ($R$ or $L$) indices during
scattering. 
The convention is that $1=$back scattering, $2=$ inter-branch forward 
scattering, $3=$ umklapp scattering, and $4=$ intra-branch forward 
scattering. Moreover, $s$ and $a$ in the subscript of $g$ characterize the spin 
dependence of the scattering matrix elements.\cite{note11}
 
The goal of the RG calculation below is to progressively eliminate high-energy electronic excitations, and study its effects on the remaining lower-energy ones. In this work the mode elimination is implemented {\it perturbatively}. To be more precise, we integrate out those $\psi_{ij\s}(k)$'s $(i=L/R, j=+/-)$ where $E_c-dE<v_F\mid k\mid <E_c$ within one-loop approximation.\cite{shankar} (Here $E_c$ is a microscopic energy cut off.)
Since the calculation is perturbative in nature, {\it the validity of the results relies on the smallness of the $g$'s}. Straightforward calculations show that among the 18 coupling constants only 12 renormalize. The resulting recursion relations are given by 
\eqa
\frac{dg^2_{4s}}{dl}&=&\frac{1}{2}(-g^1_{4s}g^1_{4s}-g^1_{4a}g^1_{4a}+
g^2_{1s}g^2_{1s}+g^2_{1a}g^2_{1a}),\nonumber \\
\frac{dg^2_{4a}}{dl}&=&g^2_{1s}g^2_{1a}-g^1_{4s}g^1_{4a},
\nonumber \\
\frac{dg^1_{4s}}{dl}&=&g^1_{4s}g^2_{4a}-g^2_{4a}g^1_{4a}-
g^1_{4s}g^1_{4s}+g^2_{1s}g^1_{2s}+g^2_{1a}g^1_{2s}\nonumber \\
&-&g^1_{2s}g^1_{2s},
\nonumber \\
\frac{dg^1_{4a}}{dl}&=&g^1_{4a}g^2_{4a}-g^2_{4a}g^1_{4s}-
g^1_{4a}g^1_{4a}+g^2_{1s}g^1_{2a}+g^2_{1a}g^1_{2a}\nonumber \\
&-&g^1_{2a}g^1_{2a},
\nonumber \\
\frac{dg^2_{1s}}{dl}&=&g^2_{4s}g^2_{1s}+g^2_{4a}g^2_{1a}-
g^1_{2s}g^1_{1s}-g^1_{2a}g^1_{1a}-g^2_{1s}g^2_{2s}\nonumber \\
&-&g^2_{1a}g^2_{2a},
\nonumber \\
\frac{dg^2_{1a}}{dl}&=&g^2_{4s}g^2_{1a}+g^2_{4a}g^2_{1s}-
g^1_{2s}g^1_{1a}-g^1_{2a}g^1_{1s}-g^2_{1s}g^2_{2a}\nonumber \\
&-&g^2_{1a}g^2_{2s},
\nonumber \\
\frac{dg^1_{2s}}{dl}&=&g^2_{4s}g^1_{2s}+g^2_{4a}g^1_{2s}+
g^1_{4s}g^2_{1s}+g^1_{4s}g^2_{1a}-2g^1_{4s}g^1_{2s}\nonumber \\
&-&g^2_{1s}g^1_{1s}
-g^2_{1a}g^1_{1a}-g^1_{2s}g^2_{2s}-g^1_{2a}g^2_{2a}, 
\nonumber \\
\frac{dg^1_{2a}}{dl}&=&g^2_{4s}g^1_{2a}+g^2_{4a}g^1_{2a}+
g^1_{4a}g^2_{1s}+g^1_{4a}g^2_{1a}-2g^1_{4a}g^1_{2a}\nonumber \\
&-&g^2_{1a}g^1_{1s}
-g^2_{1s}g^1_{1a}-g^1_{2s}g^2_{2a}-g^1_{2a}g^2_{2s}, 
\nonumber \\
\frac{dg^1_{1s}}{dl}&=&-g^1_{1s}g^1_{1s}+g^1_{1s}g^2_{2a}-
g^2_{1s}g^1_{2s}-g^2_{1a}g^1_{2a}-g^1_{1a}g^2_{2a},\nonumber \\
\frac{dg^1_{1a}}{dl}&=&-g^1_{1a}g^1_{1a}+g^1_{1a}g^2_{2a}-
g^2_{1s}g^1_{2a}-g^2_{1a}g^1_{2s}-g^1_{1s}g^2_{2a},\nonumber \\
\frac{dg^2_{2s}}{dl}&=&\frac{1}{2}(-g^2_{1s}g^2_{1s}-g^2_{1a}g^2_{1a}
-g^1_{2s}g^1_{2s}-g^1_{2a}g^1_{2a}-g^1_{1s}g^1_{1s}\nonumber \\
&-&g^1_{1a}g^1_{1a}), 
\nonumber \\
\frac{dg^2_{2a}}{dl}&=&-g^2_{1s}g^2_{1a}-g^1_{2s}g^1_{2a}
-g^1_{1s}g^1_{1a}.
\label{rec}
\eea
In the above $dl\equiv\frac{1}{\pi v_F}\ln(\frac{E_c}{E})$, where $E$ is the running energy cutoff.
% The $g$'s which do not appear in Eq.(\ref{rec}) do not %renormalize (i.e. remain unchanged).
In order to apply these recursion relations we need to determine the initial values of $g$'s. It turns out that the latter strongly depend on the topological type of the stripe. Let us concentrate on the case of small doping where $k_{FR+}\approx k_{FR-}$ and $k_{FL+}\approx k_{FL-}$.

For the topological stripe ( $\eta=1$ ) we have 
\eqa
&&g^2_{4s}=g^1_{1s}=g^2_{2s}=U, \nonumber \\
-&&g^2_{4a}= g^1_{1a}= g^2_{2a}=4a^2b^2U, \nonumber \\  &&g^1_{4a}=g^1_{2a}=g^2_{1a}=0, \nonumber \\ &&g^2_{1s}=g^1_{2s}=g^1_{4s}=(1-4a^2b^2)U.
\label{top}
\eea
For the non-topological stripe ( $\eta=-1$ ) we have
instead 
\eqa
&&g^2_{4s}=g^2_{1s}=g^2_{2s}=U, \nonumber \\ 2&&g^2_{4a}=g^1_{4a}=2g^2_{1a}=g^1_{2a}=g^1_{1a}
=2g^2_{2a}=4a^2b^2U, \nonumber \\
&&g^1_{1s}=g^1_{2s}=g^1_{4s}=(1-2a^2b^2)U.
\label{non}
\eea
In Eqs.(\ref{top}) and (\ref{non})
\eqa
&&a=r/\sqrt{1+r^2},\hspace{0.2in} b=1/\sqrt{1+r^2},\nonumber \\
&&r=\frac{M/t}{\sqrt{(M/t)^2+\Delta^2}-\Delta}.
\label{for}
\eea
In Eq.(\ref{for}) $\Delta=2\cos(k_{F0})$ for $\eta=1$, and $\Delta=1-2cos(k_{F0})$ for $\eta=-1$ respectively.\cite{note12}

%Once we set the values for initial coupling constants,
Given Eqs.(\ref{top}) and (\ref{non}), we numerically iterate Eq.(\ref{rec}) to determine the renormalized $g$'s. The result is trustworthy when i) $U/t<<1$ and ii) all renormalized $g$'s are $<<1$. Under i) and ii) we find that
in all cases there exist more than one $g$ that grow upon renormalization. To deduce the implication of these growing coupling, we compute physical susceptibilities. 

For $U=0$, the following 12 
susceptibilities capture all logarithmically divergent ones up to $R\leftrightarrow L$ and/or $+\leftrightarrow -$ exchanges: 
\eq
\chi_{\alpha}(\w)=\int dxdt e^{i\w t}<T[O_{\alpha}(x,t)O^+_{\alpha}(0,0)]>,
\ee 
where
\eqa
&&O_{cdw1}(x)=\psi_{R+\uparrow}^\dagger(x)\psi_{R-\uparrow}(x)+ \psi_{R+\downarrow}^\dagger(x)\psi_{R-\downarrow}(x), \nonumber \\
&&O_{sdw1\parallel}(x)=[\psi_{R+\uparrow}^\dagger(x)\psi_{R-\uparrow}(x)-\psi_{R+\downarrow}^\dagger(x)\psi_{R-\downarrow}(x)]/2,\nonumber \\
&&O_{sdw1\perp}(x)=[\psi_{R+\uparrow}^\dagger(x)\psi_{R-\downarrow}(x)+\psi_{R+\downarrow}^\dagger(x)\psi_{R-\uparrow}(x)]/2,\nonumber \\
&&O_{ss1}(x)= [\psi_{R+\uparrow}^\dagger(x) \psi_{R-\downarrow}^\dagger(x) -\psi_{R+ \downarrow}^\dagger(x) \psi_{R-\uparrow}^\dagger(x)]/ \sqrt{2} , \nonumber \\ 
&&O_{ts1\parallel}(x)= [
\psi_{R+\uparrow}^\dagger(x)\psi_{R-\downarrow}^\dagger(x)+\psi_{R+\downarrow}^\dagger(x)\psi_{R-\uparrow}^\dagger(x)]/ \sqrt{2}, \nonumber \\
&&O_{ts1\perp}(x)=\psi_{R+\uparrow}^\dagger(x)\psi_{R-\uparrow}^\dagger(x), \nonumber \\
&&O_{cdw2}(x)=\psi_{R+\uparrow}^\dagger(x)\psi_{L-\uparrow}(x)+\psi_{R+\downarrow}^\dagger(x)\psi_{L-\downarrow}(x), \nonumber \\
&&O_{sdw2\parallel}(x)=[\psi_{R+\uparrow}^\dagger(x)\psi_{L-\uparrow}(x)-\psi_{R+\downarrow}^\dagger(x)\psi_{L-\downarrow}(x)]/2, \nonumber \\
&&O_{sdw2\perp}(x)=[\psi_{R+\uparrow}^\dagger(x)\psi_{L-\downarrow}(x)+\psi_{R+\downarrow}^\dagger(x)\psi_{L-\uparrow}(x)]/2, \nonumber \\
&&O_{ss2}(x)= [\psi_{R+\uparrow}^\dagger(x) \psi_{L-\downarrow}^\dagger(x)- \psi_{R+\downarrow}^\dagger(x)\psi_{L-\uparrow}^\dagger(x) ] /\sqrt{2}, \nonumber \\
&&O_{ts2\parallel}(x)= [ \psi_{R+\uparrow}^\dagger(x)\psi_{L-\downarrow}^\dagger(x)+\psi_{R+\downarrow}^\dagger(x)\psi_{L-\uparrow}^\dagger(x)]/ \sqrt{2}, \nonumber \\
&&O_{ts2\perp}(x)= \psi_{R+\uparrow}^\dagger(x)\psi_{L-\uparrow}^\dagger(x). \nonumber \\
\eea 
In the above $sdw, sdw$ label the charge and spin density wave 
susceptibilities, while  $ss$ and $ts$ label the superconducting ones.
To the lowest order in $g$'s, the results are
\eq
\chi_{\ap}(\w)=A_{\ap}\chi_0(\omega)(\frac{E_c}{\omega})^{\frac{x_{\ap}}{2\pi v_F}},\hspace{0.2in}\chi_0=\frac{1}{2\pi}\ln(\frac{E_c}{\omega}).
\label{sus}
\ee
In the above the amplitudes $A$ are given by
$A_{cdw1}=A_{cdw2}=2$ and $A_{\ap}=1$ for all other $\ap$'s, and the 
exponents $x_{\ap}$ are given by
\eqa
&&x_{cdw1}=g^2_{4s}+g^2_{4a}-2g^1_{4s}, \nonumber \\
&&x_{sdw1\parallel}=  g^2_{4s}+g^2_{4a}-2g^1_{4a},\nonumber \\
&&x_{sdw1\perp}= g^2_{4s}-g^2_{4a},\nonumber \\
&&x_{ss1}=-g^2_{4s}-g^1_{4s}+g^2_{4a}+g^1_{4a},\nonumber \\
&&x_{ts1\parallel}=-g^2_{4s}+g^1_{4s}+g^2_{4a}-g^1_{4a},\nonumber \\ 
&&x_{ts1\perp}=-g^2_{4s}+g^1_{4s}-g^2_{4a}+g^1_{4a},\nonumber \\ 
&&x_{cdw2}=g^2_{2s}-2g^1_{1s}+g^2_{2a},\nonumber \\ 
&&x_{sdw2\parallel}=g^2_{2s}-2g^1_{1a}+g^2_{2a},\nonumber \\ 
&&x_{sdw2\perp}=g^2_{2s}-g^2_{2a},\nonumber \\ 
&&x_{ss2}=- g^2_{2s}-g^1_{1s}+ g^2_{2a}+g^1_{1a},\nonumber \\ 
&&x_{ts2\parallel}=g^1_{1s}-g^2_{2s}-g^1_{1a}+g^2_{2a},\nonumber \\ 
&&x_{ts2\perp}=g^1_{1s}-g^2_{2s}+g^1_{1a}-g^2_{2a}.
\label{expo}
\eea

By substituting the renormalized $g$'s into Eqs.(\ref{sus}) and (\ref{expo}) we can calculate 
$\chi_{\ap}(\w)$ as $\w\rightarrow 0$. In the following we shall restrict $M/t<<1$. We find that,
like the isolated ladder,  $\chi_{ss2}$ 
is the most divergent susceptibility for
$\eta=1$. For $\eta=-1$, we find that $\chi_{cdw1}$ and $\chi_{sdw1\perp}$ are the most 
divergent while $\chi_{ss2}$ is non-divergent.

By setting $g^j_{ia}=0$ we reproduce the results for the isolated ladder.\cite{ladder} After a comparison, we find that for $\eta=1$ superconductivity is {\it enhanced} by the non-zero $g^j_{ia}$. In other words, superconductivity is enhanced in the topological stripe!

Before proceeding to the symmetry of the superconducting order parameter, we comment on the two conditions under which the above results are obtained, namely, $M/t<<1$ and small doping. i) The dependence on $M/t$: for $\eta=1$ we find that superconductivity is replaced by spin density wave ($SDW1\perp$) when $M/t$ exceeds $\approx 0.4$. On the other hand, for $\eta=-1$, $\chi_{cdw1}$ and $\chi_{sdw1\perp}$ reamins as the most divergent susceptibilities for all values of $M/t$. ii) The dependence on doping: for large doping, the symmetry of $+\leftrightarrow -$ is lost. 
%This is true even if we assume the same Fermi velocity for %all branches. 
In that case the number of independent vertex function greatly increases. We have not done the RG calculation for the general case. What we say is that for a fixed $M\ne 0$ the effect of environment diminishes upon doping. Thus at large doping the results should resemble that of the isolated ladder. For the latter it is found that superconductivity survives for doping up to $\approx 50\%$.  

\vspace{0.1in}
\noindent{\bf{Symmetry of the superconducting order parameter}}
\vspace{0.1in}

\rm

As $\chi_{SS2}$ diverges the following operators tend to develop expectation values:
\eqa
&&b^+_1(x)= \psi_{R+\uparrow}^\dagger(x) \psi_{L-\downarrow}^\dagger(x)- \psi_{R+\downarrow}^\dagger(x)\psi_{L-\uparrow}^\dagger(x)  , \nonumber \\ 
&&b^+_2(x)= \psi_{R-\uparrow}^\dagger(x) \psi_{L+\downarrow}^\dagger(x)- \psi_{R-\downarrow}^\dagger(x)\psi_{L+\uparrow}^\dagger(x).
\label{boson} 
\eea
A straightforward mean-field analysis\cite{kll} indicates that the out-of-phase combination of $b^+_1$ and $b^+_2$, i.e., 
\eq
B^+\equiv\int dx [b^+_1(x)-b^+_2(x)],
\ee
acquires non-zero expectation. Of course, after including phase fluctuation, we expect $<B^+(x)B(x')>$ to have only quasi-off-diagonal long-range order even at $T=0$.

$b_1$ and $b_2$ might appear to be singlet Cooper pair annihilation operators. In fact this is not so. Indeed, due to the term proportional to $M$ in Eq.(\ref{str}), spin up and spin down electrons experience different one-body 
potentials. Consequently, $\psi_{R+\uparrow}^+$ and $\psi_{R+\downarrow}^+$  differ by more than a spin flip. As a result, the Cooper pair created by $B^+$ has both singlet and triplet characters. To see that more clearly, we express  $\psi^+_{R\pm\s}(k)$ and
$\psi^+_{L\pm\s}(k)$ in terms of the band operators at $M=0$: 
\eqa
&&\psi^+_{R+\s}(k)=ac^+_{a\s}(k_{FR+}+k)+\s bc^+_{a\s}( k_{FR+}+k-\pi),\nonumber \\
&&\psi^+_{R-\s}(k)=ac^+_{s\s}(k_{FR-}+k-\pi)-\s b c^+_{s\s}(k_{FR-}+k),\nonumber \\
&&\psi^+_{L+\s}(k)=ac^+_{s\s}(k_{FL+}+k+\pi)-\s b c^+_{s\s}( k_{FL+}+k),\nonumber \\
&&\psi^+_{L-\s}(k)=ac^+_{a\s}(k_{FL-}+k)+\s b c^+_{a\s}( k_{FL-}+k+\pi).
\eea
In the above $c^+_{s\s}$ and  $c^+_{a\s}$ create an electron in the symmetric and antisymmetric band respectively. Substituting the above results into Eq.(\ref{boson}) we obtain
\eqa
b^+_1&&=\int\frac{dk}{2\pi}\{a^2[c^+_{a\uparrow}(p(k))c^+_{a\downarrow}(-p(k))-(\uparrow\leftrightarrow\downarrow)] \nonumber \\&&-b^2[c^+_{a\uparrow}(p(k)+\pi)c^+_{a\downarrow}(-p(k)-\pi)-(\uparrow\leftrightarrow\downarrow)]\nonumber \\
&&+ab[c^+_{a\uparrow}(p(k)+\pi)c^+_{a\downarrow}(-p(k))+(\uparrow\leftrightarrow\downarrow)]\nonumber \\
&&-ab[c^+_{a\uparrow}(p(k))c^+_{a\downarrow}(-p(k)-\pi)+(\uparrow\leftrightarrow\downarrow)]\},\nonumber \\
b^+_2&&=\int\frac{dk}{2\pi}\{a^2[c^+_{s\uparrow}(q(k)+\pi)c^+_{s\downarrow}(-q(k)-\pi)
-(\uparrow\leftrightarrow\downarrow)] \nonumber \\&&-b^2[c^+_{s\uparrow}(q(k))c^+_{s\downarrow}(-q(k))-(\uparrow\leftrightarrow\downarrow)]\nonumber \\
&&-ab[c^+_{s\uparrow}(q(k))c^+_{s\downarrow}(-q(k)-\pi)+(\uparrow\leftrightarrow\downarrow)]\nonumber \\
&&+ab[c^+_{s\uparrow}(q(k)+\pi)c^+_{s\downarrow}(-q(k))+(\uparrow\leftrightarrow\downarrow)]\}.
\label{bb}
\eea
In the above $p(k)\equiv k_{FR+}+k$ and $q(k)\equiv k_{FR-}+k$. In Eq.(\ref{bb}) the terms proportional to $a^2$ and $b^2$ have $S=0$ and $\rm{momentum}=0$, while those proportional to $ab$ have $S=1$ and $\rm{momentum}=\pi$.

\vspace{0.1in}
\noindent{\bf{A Ginzburg-Landau theory for the mixed  symmetry}}
\vspace{0.1in}

\rm

The result obtained above for the mixed component order parameter can be understood on the basis of more general symmetry considerations. For a system having low energy antiferromagnetic and superconducting modes, the following term is allowed in the effective action:
\eq
S_{mix}=\lambda\int dt d^dx\vec{N}(x,t)\cdot[\vec{\psi}_{t}(x,t) \psi^*(x,t)+ c.c.]. 
\label{gl}
\ee

In Eq.(\ref{gl}), $\vec{N}$ is the Neel order parameter, %$\vec{M}$ is the net local magnetization,  
$\psi$ is a zero-momentum spin singlet Cooper pair field, and 
%$\vec{\psi}_{t}$-zero momentum spin triplet Cooper pair
$\vec{\psi}_{t}$ is a momentum $Q=(\pi,...,\pi)$, spin triplet Cooper-pair field.
In addition to satisfying the translational symmetry, the fact that the triplet Cooper field has a center-of-mass momentum $Q$ is crucial for Eq.(\ref{gl}) to be allowed by the point group symmetry.
Let us assume that 
\eqa
&&\int d^dx\psi^+(x)=\sum_k\phi(k)[c^+_{k\uparrow}c^+_{-k\downarrow}-c^+_{k\downarrow}c^+_{-k\uparrow}],\nonumber \\
&&\int d^dx ({\psi}^+_t)_x=\sum_k\chi(k)[c^+_{k+Q\uparrow}c^+_{-k\uparrow}+c^+_{k+Q\downarrow}c^+_{-k\downarrow}],\nonumber \\
&&\int d^dx({\psi}^+_t)_y=\sum_k\chi(k)[c^+_{k+Q\uparrow}c^+_{-k\uparrow}-c^+_{k+Q\downarrow}c^+_{-k\downarrow}]/i,\nonumber \\
&&\int d^dx({\psi}^+_t)_z=\sum_k\chi(k)[c^+_{k+Q\uparrow}c^+_{-k\downarrow}+c^+_{k+Q\downarrow}c^+_{-k\uparrow}],
\eea
where
\eqa 
&&\phi(k)=\phi(-k),\nonumber \\
&&\chi(k)=-\chi(-k-Q).
\label{x1}
\eea
The fact that $\vec{N}$ transform as the identity
under the action of the point group about a lattice site constrains the symmetry of $\phi(k)$ and $\chi(k)$. In two dimensions, where the point group is Abelian, $\phi(k)$ and $\chi(k)$ must transform identically:
\eq
\chi(k)\propto\phi(k).
\label{x2}
\ee
Putting Eqs.(\ref{x1}) and (\ref{x2}) together we obtain
\eq
\phi(k)=-\phi(k+Q). 
\label{x3}
\ee
This can be satisfied, e.g., by either for $d_{x^2-y^2}$ pairing, where
$\phi(k)\propto cos(k_x)-cos(k_y)$, or for anisotropic s-wave: $\phi(k)\propto cos(k_x)+cos(k_y)$.

%The term with net magnetization will  be present in the %external magnetic field or in disordered phase.
Depending which one among $\{\vec{N}, \vec{\psi}_t,\psi\}$ acquires expectation value, 
$S_{mix}$ causes mixing between the remaining two. For example, if $<\vec{N}>\ne 0$, $S_{mix}$ mixes $\psi$ and $\vec{\psi}_t$. This is precisely what was pointed out by Schrieffer, Wen and Zhang \cite{swz} and is a result of our calculations. 
As another example, if $<\psi>\ne 0$, then 
$\vec{N}$ hybridizes with $\vec{\psi}_t$. As a consequence, 
after integrating out $\vec{\psi}_t$, the Greens function of $\vec{N}$ is modified so that
\eq
G^R_{nn}(\vec{q},\w)=\frac{G_{nn}(\vec{q},\w)}{1-\lambda^2\mid\psi\mid^2G_{nn}(\vec{q},\w)G_{tt}(\vec{q},\w)}. 
\label{gnn}
\ee
In the above $G_{nn}$ and $G_{tt}$ are the 
Greens functions associated with $\vec{N}$ and $\vec{\psi}_t$ respectively. A resonance appears when 
\eq
1=\lambda^2\mid\psi\mid^2Re[G_{nn}(\vec{q},\w)G_{tt}(\vec{q},\w)].
\label{res}
\ee
Recently Demler and Zhang\cite{zhang} postulated that $G_{tt}$ exhibits an isolated pole, and that the
solution of Eq.(\ref{res}) near this pole is 
the observed resonance in the neutron scattering
of several high $T_c$ oxides.\cite{41} We emphasize that i) the hybridization between $\vec{N}$ and $\vec{\psi}_t$ predicted by Eq.(\ref{gl}) is generic. It does not rely on an enlarged symmetry group for the order parameters.\cite{zhang} 
%ii) Even when $G_{tt}$ does not posses an isolated pole %(hence no $SO(5)$ symmetry), a solution of Eq.(\ref{res}) %could still exist.
ii) The resonance in $G^R_{nn}$ can also appear near the resonance in $G_{nn}$. In that case the mode is predominately magnetic in origin. This would explain the 
facts that the observed cross section and the $Cu-Cu$ bilayer modulation of the resonant mode are very close to those of antiferromagnetic spin waves in the undoped antiferromagnet.\cite{ekz1} 
iii) The coupling described by Eq.(\ref{gl}) exists in the normal phase as well. As before, this coupling modifies $G_{nn}$. Since $G_{ss}$ is no longer of the form
$\mid\psi\mid^2\delta(\w)\delta(\vec{q})$ above $T_c$, it can cause the broadening of the resonance.\cite{ekz1} Of course in the latter circumstance, quasiparticle excitations will also contribute to this brodening.
%\begin{equation}
%G_{nn}\ra\frac{G_{nn}}{1-\lambda^2 G_{ss}G_{nn}G_{tt}}
%\end{equation}

%In the case that the uniform magnetization $\vec{M}$ is %important (for example when there is an external magnetic %field) we should add an addition term to Eq.(\ref{gl}):
%\eq
%S_{mix}=\lambda^{\prime}
%\int dt d^dx\vec{M}(x,t)\cdot[\vec{\psi}_t(x,t) %\psi^*(x,t)+ c.c.].
%\label{gl1}
%\ee
%In this case $\vec{M}$ mixes $\psi$ - the singlet, zero- %momentum Cooper field, with $\vec{\psi}_t$ - the triplet, %{\it zero-momentum} Cooper field. 

\vspace{0.1in}
\noindent{\bf{Some Concluding Remarks}}
\rm
\vspace{0.1in}

According to Tranquada {\it et al}, superconducting and stripe order coexist for $La_{1.6-x}Nd_{0.4}Sr_xCuO_4$ with $x=0.12,0.15,0.20$. It is tempting to relate a single stripe in such system to the model we considered above. There are two immediate worries in drawing such a connection. The first is that a real stripe is not 
necessarily made up of a pair of chains. The second is that a real stripe will not be absolutely straight. 
%The  last worry concerns about the legitimacy 
%of assuming the magnetic order in the intervening regions %being static.

We do not think the width of the stripe (as long as it is reasonably small) will qualitatively change our conclusions. The basis for that belief is that it is known for the $n$-leg Hubbard ladders that {\it away from half filling} the value of $n$ does not affect the fact that the system is superconducting for small doping.\cite{lin} 
%As to whether the magnetic order is static, we  quote %experimental findings of Aeppli {\it et al}\cite{aeppli} %that the measured magnetic structure factor $S(\vec{q},\w)$ %exhibits a quasi-elastic component below certain energy.

Now we come to the shape of stripes. When the stripes meander, the potential seeing by the electrons is no longer invariant under the translation by two lattice spacing parallel to the stripe. The irregularity presents itself as both scalar and magnetic impurities. They both have pair-breaking effect on the superconductivity as should be the case for a ``d-wave'' order parameter. Nonetheless, the basic fact that local Neel order mixes singlet with triplet  remains true.   
  
No comparison with real systems can be made without
understanding the effects of coupling between stripes. In the absence of such coupling, superconducting long-range order is not possible. Knowing how Cooper pairs tunnel from one stripe to another is crucial for the understanding of, e.g., the experimental $T_c$ versus $\delta q$ relation (here $\delta q$ is the incommensurability in neutron peak). Presumably tunneling is most efficient when two stripes ``collide'' with each other. If one assume that the average distance (along the stripe) between the collision points scales with the average distance between the stripe, then $T_c$ will scale linearly with $\delta q$ as experimentally observed.
  
%Finally, we would like to comment on the relation between
 %stripe order and superconductivity. 
In a recent paper, Emery, Kivelson and Zachar\cite{ekz} attribute the superconductivity in high $T_c$ compounds to the ``spin gap proximity effect''. In that mechanism, due to the tunneling of a pair of electrons from {\it a single metallic chain} to the environment and back, a spin gap is induced on the former. In that case, because of the spin-charge decoupling in the chain, the charge sliding mode is left as the only low energy degree of freedom. In Ref. \cite{ekz} this is identified as superconductivity.
The picture emerges from our study differs somewhat from the above. The biggest differnece is that the environment is not the cause of superconductivity. To be more specific in our model i) The magnetic environment is not spin gapped. ( We note that experimentally the quasistatic spin peak has been observed in the "striped'' compounds in $La_{1.6-x}Nd_{0.4}Sr_xCuO_4$ \cite{tranq2}.) 
ii) The stripe is superconducting even if the coupling to the environment is switched off. We should, however, mention that the coupling to the enviroment did enhance superconductivity. 
Finally, we point out that in this work  we have not addressed the issue of long-range Coulomb interaction on the superconductivity in stripes.  

Acknowledgment: We thank S. Bahcall and S.A. Kivelson for useful discussions. A.V.B. acknowledges support from the US Department of Energy.

\newpage
\bibliographystyle{unsrt}

\begin{thebibliography}{99}

\bibitem{tranq1} J.M.Tranquada {\it et al.}, Phys. Rev. Lett. {\bf73}, 1003 (1994); Nature {\bf375}, 561 (1995); Phys. Rev. B {\bf52}, 3581 (1995); Phys. Rev. B {\bf54}, 7489 (1996).

\bibitem{tranq2} J.M.Tranquada {\it et al.}, Phys. Rev. Lett. {\bf78}, 338 (1997). 


\bibitem{sep} D.Poilblanc and T.M.Rice, Phys. Rev. B{\bf39}, 9749 (1989); J.Zaanen and Gunnarson, Phys. Rev. B{\bf40}, 7391 (1989); V.J.Emery, S.Kivelson, and H.-Q.Lin, Phys. Rev. Lett.{\bf64}, 475 (1990); 
H.J.Schulz, Phys. Rev. Lett. {\bf64}, 1445 (1990); M.Kato, K.Machida, H.Nakanishi, and M.Fujita, J.Phys.Soc.Jpn. {\bf59}, 1047 (1990); T.Giamarchi and C.Lhuillier, Phys. Rev. B{\bf43}, 12943 (1991); M.Inui and P.B.Littlewood, Phys. Rev. B{\bf44}, 4415 (1991); G.An and J.M.J. van Leeuwen, Phys. Rev. B{\bf44}, 9410 (1991); J.A.Verges, F.Guinea, and E.Louis, Phys. Rev. B{\bf46}, 3562 (1992); V.J.Emery and S.Kivelson , Physica C{\bf209}, 597 (1993); P.Prelovsek and X.Zotos, Phys. Rev. B{\bf47}, 5984 (1993); P.Prelovsek and I.Sega, Phys. Rev. B{\bf49}, 15241 (1994); J.M. van Bemmel, D.F.B. ten Haaf, W. van Saarloos, J.M.J. van Leeuwen, and G.An, Phys. Rev. Lett.{\bf72}, 2442 (1994). 

\bibitem{ekz} V.J. Emery, S.A. Kivelson and O. Zachar, cond-mat 9610094.

\bibitem{cn} A.H.Castro Neto, cond-mat 9611146; cond-mat 9702180.

\bibitem{scala} S.R.White and D.J.Scalapino, cond-mat 9608138; cond-mat 9610104.

\bibitem{ladder} C.M. Varma and A. Zawadowski, Phys. Rev. B{\bf32}, 7399 (1985); 
%E. Dagotto, J. Riera and D. Scalapino, Phys. Rev. B %{\bf45}, 5744 (1992); 
%T.M. Rice, S. Gopalan and M. Sigrist, Europhys. Lett. %{\bf23}, 445 (1993); R.S. Eccleston {\it et al}, Phys. Rev. %Lett. {\bf73}, 2626 (1994); 
%M. Azuma {\it et al}, Phys. Rev. Lett. {\bf73}, 3463 %(1994); D.G. Clarke, S.P. Strong and P.W. Anderson, Phys. %Rev. Lett. {\bf72}, 3218 (1994); 
A.M. Finkelstein and A.I. Larkin, Phys. Rev. B {\bf47}, 10461 (1993);
D.V. Khveshchenko and T.M. Rice, Phys. Rev. B {\bf50}, 252 (1994); 
%K. Kuroki and H. Aoki, Phys. Rev. Lett. {\bf72}, 2947 %(1994);
%N. Nagaosa, Solid State Commun. {\bf94}, 495 (1995);
L. Balents abd M.P.A. Fisher, Phys. Rev. B {\bf53}, 12133, (1996); H.J.Shulz,  Phys. Rev. B {\bf53}, R2959, (1996).


\bibitem{note11} In the literature, (See, e.g., J. Solyom, Adv. Phys. {\bf28}, 201 (1979).) $g_{\parallel}$ and $g_{\perp}$ are used to distinguish the matrix elements when the two electrons that are scattered have the same and opposite spins
respectively. The relations between $g_{\parallel}, g_{\perp}$ and 
$g_s, g_a$ are $g_{\parallel}=g_s+g_a$, and $g_{\perp}=g_s-g_a$.

\bibitem {shankar} R.Shankar, Rev.Mod.Phys. {\bf66}, 129 (1994).

\bibitem{note12} In order to maintain the $+\leftrightarrow -$ symmetry we have evaluated the initial 
$g$'s assuming that the momentum of each incoming electron
to be either $k_{F0}$ or $-k_{F0}$. This is clearly a small-doping aprroximation.


\bibitem{kll} Yu.A.Krotov, D.-H.Lee and S.G.Louie, cond-mat 9611073.

\bibitem{swz} J.R. Schrieffer, X.-G. Wen and S.-C. Zhang,
Phys. Rev. B {\bf39}, 11663 (1989).

\bibitem{zhang} E.Demler and S.C.Zhang, Phys. Rev. Lett. {\bf75}, 4126 (1995); S-C Zhang, Science. {\bf275}, 1089 (1997).

\bibitem{ekz1} A somewhat related idea has been discussed in 
V.J. Emery, S.A. Kivelson and O. Zachar, cond-mat 9703211.
  
 
\bibitem{41} J.Rossat-Mignod {\it et al.}, Physica C {\bf185}, 86 (1991); H.Mook {\it et al.}, Phys. Rev. Lett. {\bf70}, 3490 (1994); H.F.Fong {\it et al.}, Phys. Rev. Lett. {\bf75}, 316 (1995); {\bf78}, 713 (1997); P.Dai {\it et al.}, Phys. Rev. Lett. {\bf77}, 5425 (1996). 
 
\bibitem{lin} H-H Lin, L. Balents and M.P.A. Fisher, Cond-mat/9703055.

%\bibitem{aeppli} T.E.Mason, G.Aeppli, and H.A.Mook {\it et %al.}, Phys. Rev. Lett. {\bf68}, 1414 (1992); S.M.Hayden, %T.E.Mason, G.Aeppli, H.A.Mook, S.-W.Cheong, and Z.Fisk, in %{\it Phase Separation in Cuprate Superconductors} (World %Scientific, Singapore, 1993). 
\end{thebibliography}

\vspace*{\fill}
\end{document}